\documentclass[amsmath,amssymb,aps,prd,twocolumn,groupedaddress]{revtex4-1}
\begin{document}


\title{Second order perturbations of relativistic membranes in curved spacetime}


\author{V. Kiosses}
\email[]{bkios@physics.auth.gr}
\author{A. Nicolaidis}
\email[]{nicolaid@auth.gr}
\affiliation{Theoretical Physics Department\\Aristotle University of Thessaloniki\\
54124 Thessaloniki, Greece}


\date{\today}

\begin{abstract}
 A manifestly covariant equation is derived to describe the second order perturbations in topological
 defects and membranes
 on arbitrary curved background spacetimes. This, on one hand, generalizes work on macroscopic strings 
 in Minkowski spacetime
 and introduces a framework for studing in a precise manner membranes behavior near the black hole horizon
 and on the other hand, introduces a more general framework for examining the stability of topological defects in
 curved spacetimes.
\end{abstract}


\maketitle

\section{Introduction}
Perturbations have been extensively used in studing the physical
properties of topological defects and relativistic membranes.
The interest in perturbative approaches has,
at least, a twofold motivation. First, topological defects of one form
or another are inevitably formed during phase transitions in the early
universe. Their cosmological implications, however, appear to depend
sensitively on their stability with respect to perturbations. Guven,
in the approximation of Garriga and Vilenkin \cite{GarVil91}, has introduced
a framework for examining the stability of topological defects moving
in a general curved background spacetime \cite{Gu93}. His approach to perturbation
theory was to expand the action describing the evolution of the defect
in a manifestly covariant way out to first order around a given classical
solution. Secondly, let us consider the motion of a membrane in curved
background spacetime. This system is endowed with conformal symmetry
on the world sheet and the membrane equations of motion are supplied
with corresponding constraints. The symmetry and the constraints underline
the fact that only the transverse membrane motion is physical. Therefore
the actual physical degrees of freedom of a membrane are less than
the dimensions of spacetime. There are two alternatives open: Either
work with all degrees of freedom and check at every stage of the calculation
that the constraints are satisfied; or work directly with the transverse
degrees of freedom. Choosing the second approach there is a price
to be paid, the membrane equations of motion become nonlinear. However,
that is not really a problem in a perturbative scheme where the equations
of motion are to be solved order by order in the expansion around
the zeroth order unperturbed membrane.

Until now only first order perturbations around the zeroth
order membrane and string in arbitrary curved background have been
considered \cite{deVSan87,FrLa94,Car93}. This is perfectly enough for many purposes, but in certain
cases it is necessary to consider also the second order perturbations.
For instance, considering small perturbations around a contracting
circular string, it is easy to see that there is no contribution to
the total conserved energy to first order; the first order contribution
simply integrates out. The first nonzero contribution (besides the
zeroth order contribution) to the total energy is quadratic in the
first order perturbations, but then also second order perturbations
must be included for consistency, since they contribute to the same
order. One of us (AN) in collaboration with A. Larsen 
managed to generalize
the results for first order perturbations to second order perturbations
but in flat backgrounds \cite{NicLar01}.

The purpose of the present paper is to derive a manifestly covariant equation to
describe the second order perturbations in a membrane on arbitrary curved backgrounds. In
order to do this we exploit the kinematical framework of Capovilla
and Guven \cite{CapGuv95} for describing deformations of an arbitrary world
sheet. This language provides us a geometrical approach to the description
of perturbations. Within this framework,
the equations of motion and both the equations of motion describing
perturbations about classical solutions and equations of motion for
the second order perturbations can be constructed in lego block fashion,
by assembling the various kinematical ingredients. 

The paper is organised as follows. To establish our notation
we begin in section II by summarizing the well-known classical kinematical
description provided by the Gauss-Weingarten equations of an embedded
timelike world sheet of dimension $n$ in a spacetime of dimension
$n+p$, in terms of its intrinsic and extrinsic geometries \cite{CapGuv95}.
In this same kinematical spirit, in section III, we describe the deformation of the
world sheet. There are analogues of the Gauss-Weingarten equations
which are useful for identifying the structures associated with such
deformations. In section IV we apply this kinematical framework to
Nambu action and reproduce the corresponding equations of motion. In the same section we
discuss the equations of motion describing first order perturbations
on arbitrary curved backgrounds and improve the perturbative expansion by
including second order terms. We conclude in
section V with a brief discussion.

\section{Mathematics of submanifolds and embeddings}

In this section we provide an overview of the well-known
mathematical description of the world sheet of a membrane viewed as
an embedded surface (submanifold) in a fixed background spacetime
(manifold).

Let $M$ be an $n+p$-dimensional manifold, $N$ an $n$-dimensional
manifold and $\theta:N\rightarrow M$ is an imbedding. 
The image $\theta\left(N\right)$ of $N$ is said to be a submanifold
of $M$. If $\left\{ x^{a}\right\} ,\; a=1,\cdots,n$, are differentiable
coordinates on the manifold $N$, then the submanifold $\theta\left(N\right)$
can be given locally by the equations 
\begin{flalign}
 &  &  &  & y^{\mu}=y^{\mu}\left(x^{a}\right), &  & \left(\mu=1,\cdots,n+p\right)\label{eq:1}
\end{flalign}
where $y^{\mu}$ are differentiable functions of
the variables $x^{a}$ and the rank of the matrix $\left(\frac{\partial y^{\mu}}{\partial x^{a}}\right)$
is equal to $n$. The $n$ vectors 
\begin{equation}
e_{a}:=\frac{\partial y^{\mu}}{\partial x^{a}}\frac{\partial}{\partial y^{\mu}}\equiv y_{,a}^{\mu}\partial_{\mu}
\end{equation}
form a basis of tangent vectors to $\theta\left(N\right)$ at each
point of $\theta\left(N\right)$. The compoments of the induced metric
on the world sheet is then given by
\begin{equation}
\gamma_{ab}=g_{\mu\nu}\frac{\partial y^{\mu}}{\partial x^{a}}\frac{\partial y^{\nu}}{\partial x^{b}}=g\left(e_{a},e_{b}\right).
\end{equation}

Consider the vector $N^{j}=n_{i}g^{ij}$, with $n_{i}$ the $i$th
unit, to be orthogonal to all the vectors tangent to $\theta\left(N\right)$.
Then if $\left\{ e_{a}\right\} ,\; a=1,\cdots,n$ are a basis of tangent
vectors to $\theta\left(N\right)$, $\left\{ e_{a},n_{i}\right\} $
(where $\left\{ n_{i}\right\} ,\; i=1,\cdots,p$,) will be linearly
independent and can be used as a basis for the spacetime vectors of
$M$. The components of $g$ with respect to this basis will be 
\begin{equation}
g_{\mu\nu}=\left(\begin{array}{cc}
g\left(n^{i},n^{j}\right) & 0\\
0 & g\left(e_{a},e_{b}\right)
\end{array}\right).
\end{equation}
As the metric $g$ is assumed to be non-degenerate, this shows that
$g\left(n^{i},n^{j}\right)=g^{\mu\nu}n_{\mu}^{i}n_{\nu}^{j}\neq0$.
So one can normalize the normal forms $\left\{ n^{k}\right\} $ to
have unit magnitude, i.e. $g^{\mu\nu}n_{\mu}^{i}n_{\nu}^{j}=\delta^{ij}$.
Thus $\left\{ n^{k}\right\} $ defined by
\begin{equation}
g\left(n^{i},n^{j}\right)=\delta^{ij},\qquad g\left(e_{a},n^{j}\right)=0.
\end{equation}

Normal vielbein indices are raised and lowered with $\delta^{ij}$
and $\delta_{ij}$, respectively, whereas tangential indices are raised
and lowered with $\gamma^{ab}$ and $\gamma_{ab}$, respectively. 

We define the world sheet
projections of the spacetime covariant derivatives with $D_{a}:=e_{a}^{\mu}D_{\mu}$,
where $D_{\mu}$ is the covariant derivative compatible with $g_{\mu\nu}$.
Let us now consider the world sheet gradients of the basis vectors
$\left\{ e_{a},n^{i}\right\} $, $D_{a}e_{b}$ and $D_{a}n^{i}$.
Since $\left(D_{a}e_{b}\right)_{x}$ and $\left(D_{a}n^{i}\right)_{x}$
are defined for each $x\in N$, we can decompose each one of these
gradients into a tangential and a normal component:
\begin{eqnarray}
\left(D_{a}e_{b}\right)_{x} & = & \left(\nabla_{a}e_{b}\right)_{x}+a_{x}\left(e_{a},e_{b}\right),\label{eq:6}\\
\left(D_{a}n^{i}\right)_{x} & = & a_{x}\left(e_{a},n^{i}\right)+\left(\mathcal{D}_{a}n^{i}\right)_{x},\label{eq:7}
\end{eqnarray}
where we denote by $\nabla_{a}$ the world sheet covariant derivative,
by $\mathcal{D}_{a}$ the world sheet covariant derivative defined
on fields transforming as tensors under normal frame rotations and
by $a_{x}$ the extrinsic curvature of the world sheet. Writing the
quantities $D_{a}e_{b}$ and $D_{a}n^{i}$ with respect to the basis
vectors $\left\{ e_{a},n^{i}\right\} $, equations (\ref{eq:6}) and
(\ref{eq:7}) become:
\begin{eqnarray}
D_{a}e_{b} & = & \gamma_{ab}^{c}e_{c}-K_{ab}^{i}n_{i},\label{eq:8}\\
D_{a}n^{i} & = & K_{ac}^{i}e^{c}+\omega_{a}^{ij}n_{j}.\label{eq:9}
\end{eqnarray}

These kinematical expressions, generalizing the classical Gauss-Weingarten
equations, describe completely the extrinsic geometry of the world
sheet.

The $\gamma_{ab}^{c}$ are the connection coefficients compatible
with the world sheet metric $\gamma_{ij}$:
\begin{equation}
\gamma_{ab}^{c}=g\left(D_{a}e_{b},e^{c}\right)=\gamma_{ba}^{c}.
\end{equation}
The quantity $K_{ab}^{i}$ is the $i$th extrinsic curvature of the
world sheet:
\begin{equation}
K_{ab}^{i}=-g\left(D_{a}e_{b},n^{i}\right)=K_{ba}^{i}.
\end{equation}
The normal fundamental form, or extrinsic twist potential, of the
world sheet is defined by 
\begin{equation}
\omega_{a}^{ij}=g\left(D_{a}n^{i},n^{j}\right)=-\omega_{a}^{ji}.
\end{equation}

Equations (\ref{eq:8}) and (\ref{eq:9}) will help us find
a relationship between the curvature tensor of spacetime $M$ and
the curvature tensor of world sheet $N$. By definition the Riemann
tensor of the spacetime covariant derivative $D_{\alpha}$ is given
by: 
\begin{equation}
R\left(e_{\alpha},e_{\beta}\right)e_{\gamma}=D_{\alpha}\left(D_{\beta}e_{\gamma}\right)-D_{\beta}\left(D_{\alpha}e_{\gamma}\right)-D_{\left[\alpha,\beta\right]}e_{\gamma}
\end{equation}
Thus, taking advantage of Gauss - Weingarten equations (\ref{eq:8}-\ref{eq:9})
we derived the Gauss - Codazzi, Codazzi - Mainardi and Ricci integrability
conditions:
\begin{eqnarray*}
g\left(R\left(e_{b},e_{a}\right)e_{c},e_{d}\right) & = & R_{abcd}-K_{ac}^{i}K_{bdi}+K_{ad}^{i}K_{bci},\\
g\left(R\left(e_{b},e_{a}\right)e_{c},n^{i}\right) & = & \mathcal{D}_{a}K_{bc}^{i}-\mathcal{D}_{b}K_{ac}^{i},\\
g\left(R\left(e_{b},e_{a}\right)n^{i},n^{j}\right) & = & \Omega_{ab}^{ij}-K_{ac}^{i}K_{b}^{cj}+K_{bc}^{i}K_{a}^{cj}.
\end{eqnarray*}
We use the notation $g\left(R\left(Y_{1},Y_{2}\right)Y_{3},Y_{4}\right)=R_{\alpha\beta\gamma\delta}Y_{2}^{\alpha}Y_{1}^{\beta}Y_{3}^{\gamma}Y_{4}^{\delta}$.
$R_{\beta\gamma\delta}^{\alpha}$ is the Riemannian curvature tensor
of spacetime, whereas $R_{bcd}^{a}$ is the Riemann tensor of the
world sheet covariant derivative $\nabla_{a}$, and $\Omega_{ab}^{ij}$
is the curvature associated with $\omega_{a}^{ij}$.

\section{Deformation in the imbedding\label{sec:4-Deformation-in-the}}

Let us now consider the neighboring $n$-dimensional surface
$N'$ described by a deformation of $N$:
\begin{equation}
y^{\mu}=y^{\mu}\left(x^{a}\right)+\delta y^{\mu}\left(x^{a}\right).\label{eq:14}
\end{equation}

According to \cite{HawEl73} this situation can be investigated
by examining the behaviour of a congruence of curves with tangent
vector $\vec{V}$. These curves could represent the histories of small
test particles, in which case they would be geodesics. Suppose $\lambda\left(t\right)$
is a curve with tangent vector $\vec{\delta}=\left(\partial/\partial t\right)_{\lambda}$.
Then one may construct a family $\lambda\left(t,s\right)$ of curves
by moving each point of the curve $\lambda\left(t\right)$ a distance
$s$ along the integral curves of $\vec{V}$. If one now defines $\vec{\delta}$
as $\left(\partial/\partial t\right)_{\lambda\left(t,s\right)}$ it
follows that the Lie brackets $\left[\vec{\delta},\vec{V}\right]$
is equal to zero (also the Lie derivative, $L_{\vec{V}}\vec{\delta}$,
is zero), so because we deal only with torsion-free connections we
take the equality
\begin{equation}
D_{\vec{\delta}}\vec{V}=D_{\vec{V}}\vec{\delta}.\label{eq:15}
\end{equation}

One may interpret $\vec{\delta}$ as representing the separation of
points equal distances from some arbitrary initial points along two
neighbouring curves. If one adds a multiple of $\vec{V}$ to $\vec{\delta}$
then this vector will represent the separation of points on the same
two curves but at different distances along the curves.

Let us choose $\lambda\left(t=0,s\right)$ as a curve belonging
to $N$, then $\lambda\left(t,s\right)$ describe a curve in the neighboring
surface $N'$. Thus we can define $\vec{\delta}=\left(\partial/\partial t\right)_{\lambda}$
as the deformation vector field ($\vec{\delta}=\delta y$) and decomposing
it with respect to the spacetime basis $\left\{ e_{a},n^{i}\right\} $,
we take
\begin{equation}
\vec{\delta}=\Phi^{a}e_{a}+\Phi^{i}n_{i}.
\end{equation}
The tangential projection can always be identified with the action
of a world-sheet diffeomorphism, $\delta^{\mu}=\Phi^{a}y_{,a}^{\mu}$,
and so will subsequently be ignored. The physically observable measure
of the deformation is therefore provided by the projection of $\vec{\delta}$
orthogonal to $N$, characterized by the $p$ scalar fields $\Phi^{i}$.

The displacement $\delta^{\mu}=\delta y^{\mu}$ in
the embedding induces a displacement in both the tangent basis $\left\{ e_{a}\right\} $and
normal basis $\left\{ n_{i}\right\} $. In light of the discussion
above, let $\vec{\delta}=\Phi^{i}n_{i}$, and consider the gradients
of $\left\{ e_{a}\right\} $ and $\left\{ n_{i}\right\} $ along the
vector field $\delta^{\mu}$, defined with $D_{\delta}$.

We can always expand $D_{\delta}e_{a}$ and $D_{\delta}n_{i}$
with respect to the spacetime basis $\left\{ e_{a},n_{i}\right\} $,
in a way analogous to the Gauss -Weingarten equations as
\begin{eqnarray}
D_{\delta}e_{a} & = & \beta_{ab}e^{b}+J_{aj}n^{j},\label{eq:17}\\
D_{\delta}n_{i} & = & -J_{ai}e^{a}+\gamma_{ij}n^{j}.\label{eq:18}
\end{eqnarray}
Comparison (\ref{eq:17}) with the Gauss equation (\ref{eq:8}) shows
that the quantity $\beta_{ab}$, defined by
\begin{equation}
\beta_{ab}=g\left(D_{\delta}e_{a},e_{b}\right)=\beta_{ba},\label{eq:19}
\end{equation}
appears in the same position as $\gamma_{ab}^{c}$. The quantities
$J_{aj}$ are defined by 
\begin{equation}
J_{aj}=g\left(D_{\delta}e_{a},n_{j}\right),\label{eq:20}
\end{equation}
and appear in the same position as $K_{ab}^{i}$ in (\ref{eq:8},\ref{eq:9}).
We note that $\beta_{ab}$ transforms as a scalar under normal frame
rotations, whereas $J_{aj}$ transforms as a vector.

The normal projection of $D_{\delta}n_{i}$,
\begin{equation}
\gamma_{ij}=g\left(D_{\delta}n_{i},n_{j}\right)=-\gamma_{ji},\label{eq:21}
\end{equation}
is a new structure we have not encountered already. In contrast to
$\beta_{ab}$ and $J_{aj}$, however, there is no simple relationship
between $\gamma_{ij}$ and deformations of the world sheet. The analogy
between (\ref{eq:18}) and (\ref{eq:9}) suggests a role for $\gamma_{ij}$
analogous to $\omega_{a}^{ij}$. For this purpose we introduce a covariant
deformation derivative as follows
\begin{equation}
\overline{D}_{\delta}\Psi_{i}=D_{\delta}\Psi_{i}+\gamma_{i}^{j}\Psi_{j}.\label{eq:22}
\end{equation}
Equation (\ref{eq:18}) can then be written in the form
\begin{equation}
\overline{D}_{\delta}n_{i}=-J_{ai}e^{a}.\label{eq:23}
\end{equation}
Using (\ref{eq:15}), it is easy to show that
\begin{align}
\beta_{ab}=g\left(D_{\delta}e_{a},e_{b}\right)=g\left(D_{a}\vec{\delta},e_{b}\right) & =g\left(D_{a}n^{i},e_{b}\right)\Phi_{i}\nonumber \\
 & =K_{ab}^{i}\Phi_{i},\label{eq:24}
\end{align}
and 
\begin{align}
J_{aj}=g\left(D_{\delta}e_{a},n_{j}\right) & =g\left(D_{a}\vec{\delta},n_{j}\right)\nonumber \\
 & =g\left(D_{a}n^{i},n^{j}\right)\Phi_{i}+\nabla_{a}\Phi_{j}\nonumber \\
 & =\omega_{a}^{ij}\Phi_{i}+\nabla_{a}\Phi_{j}\nonumber \\
 & =\mathcal{D}_{a}\Phi_{j}.\label{eq:25}
\end{align}
The deformation in the induced metric on $N$ is just twice $\beta_{ab}$:
\begin{align}
D_{\delta}\gamma_{ab}=D_{\delta}g\left(e_{a},e_{b}\right) & =2g\left(e_{a},D_{\delta}e_{b}\right)\nonumber \\
 & =2\beta_{ab}=2K_{ab}^{i}\Phi_{i}.\label{eq:26}
\end{align}

Let us now evaluate the deformation of the extrinsic curvatures,
$\overline{D}_{\delta}K_{ab}^{i}$. Using its definition we have that

\begin{equation}
\overline{D}_{\delta}K_{ab}^{i}=-g\left(\overline{D}_{\delta}n^{i},D_{a}e_{b}\right)-g\left(n^{i},D_{\delta}D_{a}e_{b}\right).\label{eq:27}
\end{equation}
Using Eq. (\ref{eq:23}) and the Gauss equation (\ref{eq:8}), the
first term on the right-hand side is given by
\begin{equation}
-g\left(\overline{D}_{\delta}n^{i},D_{a}e_{b}\right)=\gamma_{ab}^{c}J_{c}^{i}.\label{eq:28}
\end{equation}
The second term on the right-hand side can be developed using the
Ricci identity, as
\begin{widetext}
\begin{align*}
-g\left(n^{i},D_{\delta}D_{a}e_{b}\right) & =-g\left(n^{i},R\left(\vec{\delta},e_{a}\right)e_{b}\right)-g\left(n^{i},D_{a}D_{\delta}e_{b}\right)\\
 & =-g\left(n^{i},R\left(n_{j},e_{a}\right)e_{b}\right)\Phi^{j}-D_{a}J_{b}^{i}+\beta_{bc}\cdot K^{ic}{}_{a}+\omega_{a}^{ij}\cdot J_{bj}\\
 & =-g\left(n^{i},R\left(n_{j},e_{a}\right)e_{b}\right)\Phi^{j}-\mathcal{D}_{a}\mathcal{D}_{b}\Phi^{i}+K_{bcj}\cdot K^{ic}{}_{a}\cdot\Phi^{j},
\end{align*}
\end{widetext}
where in the last line we have used eqs. (\ref{eq:25}) and (\ref{eq:24}).
Therefore we find 
\begin{align}
\overline{D}_{\delta}K^{i}{}_{ab}= & -\mathcal{D}_{a}\mathcal{D}_{b}\Phi^{i}\nonumber \\
 & +\left[g\left(n^{i},R\left(n_{j},e_{a}\right)e_{b}\right)+K_{bcj}\cdot K^{ic}{}_{a}\right]\Phi^{j}.\label{eq:29}
\end{align}
Note that the change of the extrinsic curvature under an infinitesimal
deformation of the world sheet involves second derivatives of the
scalar fields $\Phi^{i}$.

\section{The Equations of motion\label{sec:5-eq. of motion}}

In this section we apply the kinematical framework we have
presented to the derivation of the equations of motion of relativistic
membranes of physical interest. The action that we are going to use
is the most simple generally covariant action one can associate with
the surface, proportional to the area swept out by $N$:
\begin{equation}
S=-\sigma\int_{N}d^{n}x\sqrt{-\gamma}.\label{eq:30}
\end{equation}
where $\gamma=det\gamma_{ab}$ and $\sigma$ is the membrane tension.
The equations of motion are given by the extrema of $S$
subject to variations
\begin{equation}
y^{\alpha}=y^{\alpha}\left(x^{m}\right)+\delta y^{\alpha}\left(x^{m}\right),\label{eq:31}
\end{equation}
 so,
\begin{align}
\delta S & =-\sigma\int_{N}d^{n}x\left(-\frac{1}{2\sqrt{-\gamma}}\delta\gamma\right)\nonumber \\
 & =-\frac{1}{2}\sigma\int_{N}d^{n}x\left(\sqrt{-\gamma}\gamma^{ab}\delta\gamma_{ab}\right)\label{eq:32}
\end{align}
 where we have used Jacobi's formula, the rule for differentiating
a determinant. Eq. (\ref{eq:26}) tells us that the quantity $\delta\gamma_{ab}$
is equal to
\begin{equation}
\delta\gamma_{ab}=2K^{i}{}_{ab}\Phi_{i},\label{eq:33}
\end{equation}
 remember we take into account only the projection of variation $\delta y^{\alpha}$
othogonal to N. Therefore from the variational principle $\delta S=0$,
the equations of motion are given by 
\begin{equation}
\gamma^{ab}\cdot K^{i}{}_{ab}=0.\label{eq:34}
\end{equation}

\subsection{First order perturbations}

In order to derive the equations of motion for first order
perturbations of a membrane, we simply have to consider
the linearization of eq.(\ref{eq:34}). This is done by writing the
perturbed quantities $\widetilde{\gamma}^{ab}=\gamma^{ab}+\delta\gamma^{ab}$
and $\widetilde{K}^{i}{}_{ab}=K^{i}{}_{ab}+\delta K^{i}{}_{ab}$ in
eq. (\ref{eq:34}) and holding the first order terms:

\begin{align}
\widetilde{\gamma}^{ab}\widetilde{K}^{i}{}_{ab} & =0\nonumber \\
\gamma^{ab}K^{i}{}_{ab}+\delta\gamma^{ab}K^{i}{}_{ab}+\gamma^{ab}\delta K^{i}{}_{ab} & =0\nonumber \\
\delta\gamma^{ab}K^{i}{}_{ab}+\gamma^{ab}\delta K^{i}{}_{ab} & =0\label{eq:35}
\end{align}
The same result we derive taking the quadratic order of action (\ref{eq:30}).
Thus we should compute the quantity $\delta\gamma^{ab}\cdot K^{i}{}_{ab}+\gamma^{ab}\cdot\delta K^{i}{}_{ab}$.
This is easy, using eqs. (\ref{eq:26}) and (\ref{eq:29}) we have:
\begin{widetext}
\begin{align*}
\delta\gamma^{ab}\cdot K^{i}{}_{ab}+\gamma^{ab}\cdot\delta K^{i}{}_{ab}= & -2K^{jab}{}K^{i}{}_{ab}\Phi_{j}-\gamma^{ab}\cdot\mathcal{D}_{a}\mathcal{D}_{b}\Phi^{i}\\
 & +\gamma^{ab}\left[g\left(n^{i},R\left(n_{j},e_{a}\right)e_{b}\right)+K_{bcj}\cdot K^{ic}{}_{a}\right]\Phi^{j}\\
= & -\triangle\Phi^{i}+g\left(n^{i},R\left(n_{j},e_{a}\right)e^{a}\right)\cdot\Phi^{j}-K_{baj}\cdot K^{iba}\cdot\Phi^{j}
\end{align*}
\end{widetext}
where
\begin{equation}
 \triangle=\gamma^{ab}\cdot\mathcal{D}_{a}\mathcal{D}_{b}
\end{equation}
and the equations of motion for the first order perturbations take the form:
\begin{equation}
\triangle\Phi^{i}-g\left(n^{i},R\left(n_{j},e_{a}\right)e^{a}\right)\cdot\Phi^{j}+K_{baj}\cdot K^{iba}\cdot\Phi^{j}=0.\label{eq:37}
\end{equation}

\subsection{Second order perturbation\label{sec:6-Second-order-perturbation}}

We continue to the derivation of the equations of motion for
the second order perturbations in world-sheet, which has not been obtained before. For this case, we write
$y^{\mu}$ in the following way:
\begin{equation}
y^{\mu}=y^{\mu}\left(x^{a}\right)+\delta y^{\mu}_{(1)}\left(x^{a}\right)+\delta y^{\mu}{}_{(2)}\left(x^{a}\right).\label{eq:38}
\end{equation}
Since we are interested only in physical (transverse) perturbations,
we have already said that $\delta y^{\mu}=\Phi^{i}n_{i}^{\mu}$.
Thus, by expanding up to second order the deformation vector field
$\vec{\delta}'$, we have:

\begin{align}
\vec{\delta}' & =(\Phi^{i}+D_{\delta}\Phi^{i})(n_{i}+D_{\delta}n_{i})\nonumber \\
 & =\Phi^{i}n_{i}+(D_{\delta}\Phi^{i})n_{i}+\Phi^{i}D_{\delta}n_{i}\nonumber \\
 & =\underbrace{\Phi^{i}n_{i}}_{\vec{\delta}^{(1)}}+\underbrace{\Psi^{i}n_{i}+\Phi^{i}D_{\delta}n_{i}}_{\vec{\delta}^{(2)}},\label{eq:39}
\end{align}
so in compare with (\ref{eq:38}) we take
\begin{equation}
\delta y^{\mu}{}_{(2)}\left(x^{a}\right)=\Psi^{i}n_{i}+\Phi^{i}D_{\delta}n_{i},
\end{equation}
with the new deformation vector to take the form:
\begin{equation}
\vec{\delta}^{(2)}=\Psi^{i}n_{i}+\Phi^{i}D_{\delta}n_{i}.\label{eq:41}
\end{equation}
In (\ref{eq:41}) there is the first order perturbation of the normal
vector, $D_{\delta}n_{i}$, which is given from (\ref{eq:18}).
We consider now the gradients of $\left\{ e_{a}\right\} $ and $\left\{ n_{i}\right\} $
along the vector field $\delta^{\mu(2)}$, defined with $D_{\delta^{(2)}}$.
We can expand $D_{\delta^{(2)}}e_{a}$ and $D_{\delta^{(2)}}n_{i}$
with respect to the spacetime basis $\left\{ e_{a},n_{i}\right\} $,
in a way analogous to (\ref{eq:17}) and (\ref{eq:18}) respectively
as
\begin{equation}
D_{\delta^{(2)}}e_{a}=\beta_{ab}^{(2)}e^{b}+J_{aj}^{(2)}n^{j},\label{eq:42}
\end{equation}
and
\begin{equation}
D_{\delta^{(2)}}n_{i}=-J_{ai}^{(2)}e^{a}+\gamma_{ij}^{(2)}n^{j}.\label{eq:43}
\end{equation}
 Quantity $\beta_{ab}^{(2)}(\neq\beta_{ab})$ is defined by the relation
\begin{equation}
\beta_{ab}^{(2)}=g(D_{\delta^{(2)}}e_{a},e_{b})=\beta_{ba}^{(2)}\label{eq:44}
\end{equation}
 and $J_{ai}^{(2)}(\neq J_{ai})$ by the relation 
\begin{equation}
J_{ai}^{(2)}=g(D_{\delta^{(2)}}e_{a},n_{i}).\label{eq:45}
\end{equation}
The quantities $\beta_{ab}^{(2)}$ and $J_{ai}^{(2)}$ can be expressed in terms of scalar fields $\Psi^{i}$ 
and $\Phi^{i}$, in a way analogous to the equations (\ref{eq:24}) and (\ref{eq:25}), considering the equation
\begin{equation}
D_{\delta^{(2)}}e_{a}=D_{a}\delta^{\mu(2)},\label{eq:46}
\end{equation}
which is the analogue for deformation vector field $\vec{\delta}'$ of equation (\ref{eq:15}). 
Using (\ref{eq:46}), we show that:
\begin{widetext}
\begin{align}
\beta_{ab}^{(2)}=g(D_{\delta^{(2)}}e_{a},e_{b})=g(D_{a}\delta^{\mu(2)},e_{b})= & g(D_{a}(\Psi^{i}n_{i}),e_{b})+g(D_{a}(\Phi^{i}D_{\delta}n_{i}),e_{b})\nonumber \\
= & g(D_{a}n_{i},e_{b})\Psi^{i}-g(D_{a}(\Phi^{i}J_{ci}e^{c}),e_{b})\nonumber \\
= & K_{ab}{}^{i}\Psi_{i}-\gamma_{b}^{c}(\nabla_{a}\Phi^{i})\cdot(\mathcal{D}_{c}\Phi_{i})\nonumber \\
 & -\gamma_{b}^{c}[\nabla_{a}(\mathcal{D}_{c}\Phi_{i})]\Phi^{i}-g(\gamma_{ad}{}^{c}e^{d}-K_{aj}{}^{c}n^{j},e_{b})\Phi^{i}J_{ci}\nonumber \\
= & K_{ab}{}^{i}\Psi_{i}-(\nabla_{a}\Phi^{i})\cdot(\mathcal{D}_{b}\Phi_{i})-[\nabla_{a}(\mathcal{D}_{b}\Phi_{i})]\Phi^{i}\nonumber \\
 & -\gamma_{b}^{d}\cdot\gamma_{ad}{}^{c}\cdot\Phi^{i}(\mathcal{D}_{c}\Phi_{i})+\omega_{ak}^{i}\Phi^{k}(\mathcal{D}_{b}\Phi_{i})-\omega_{ak}^{i}\Phi^{k}(\mathcal{D}_{b}\Phi_{i})\nonumber \\
= & K_{ab}{}^{i}\Psi_{i}-(\nabla_{a}\Phi^{i})\cdot(\mathcal{D}_{b}\Phi_{i})-\omega_{ak}^{i}\Phi^{k}(\mathcal{D}_{b}\Phi_{i})\nonumber \\
 & -[\nabla_{a}(\mathcal{D}_{b}\Phi_{i})]\Phi^{i}-\gamma_{ab}{}^{c}\cdot(\mathcal{D}_{c}\Phi_{i})\cdot\Phi^{i}-\omega_{ai}^{k}(\mathcal{D}_{b}\Phi_{k})\cdot\Phi^{i}\nonumber \\
= & K_{ab}{}^{i}\Psi_{i}-(\mathcal{D}_{a}\Phi^{i})\cdot(\mathcal{D}_{b}\Phi_{i})-[\mathcal{D}_{a}(\mathcal{D}_{b}\Phi_{i})]\Phi^{i}.\label{eq:47}
\end{align}
and
\begin{align}
J_{ai}^{(2)}=g(D_{\delta^{(2)}}e_{a},n_{i})=g(D_{a}\delta^{\mu(2)},n_{i})= & g(D_{a}(\Psi^{j}n_{j}),n_{i})+g(D_{a}(\Phi^{j}D_{\delta}n_{j}),n_{i})\nonumber \\
= & g(D_{a}n_{j},n_{i})\Psi^{j}+g((D_{a}\Psi^{j})n_{j},n_{i})-g(D_{a}(\Phi^{j}J_{cj}e^{c}),n_{i})\nonumber \\
= & \delta_{ki}\cdot\omega_{aj}{}^{k}\cdot\Psi^{j}+\nabla_{a}\Psi_{i}\nonumber \\
 & -g(\gamma_{af}{}^{c}\cdot e^{f}-K_{al}{}^{c}\cdot n^{l},n_{i})\Phi^{j}J_{cj}\nonumber \\
= & \mathcal{D}_{a}\Psi_{i}+\delta_{i}^{l}\cdot K_{al}{}^{c}J_{cj}\Phi^{j}\nonumber \\
= & \mathcal{D}_{a}\Psi_{i}+K_{ai}{}^{c}(\mathcal{D}_{c}\Phi_{j})\Phi^{j}\label{eq:48}
\end{align}
\end{widetext}

In order to derive the equations of motion for second order perturbations
of a membrane, we substitute the
perturbed quantities $\widetilde{\gamma}^{ab}=\gamma^{ab}+\delta^{(1)}\gamma^{ab}+\delta^{(2)}\gamma^{ab}$
and $\widetilde{K}^{i}{}_{ab}=K^{i}{}_{ab}+\delta^{(1)}K^{i}{}_{ab}+\delta^{(2)}K^{i}{}_{ab}$
in eq. (\ref{eq:34}) and keep the quadratic terms:
\begin{equation}
\delta^{(2)}\gamma^{ab}K^{i}{}_{ab}+\gamma^{ab}\delta^{(2)}K^{i}{}_{ab}+(\delta^{(1)}\gamma^{ab})(\delta^{(1)}K^{i}{}_{ab})=0\label{eq:49}
\end{equation}
In order to proceed we have to evaluate the quantities $\delta^{(2)}\gamma_{ab}$ and
$\delta^{(2)}K^{i}{}_{ab}$.
Let us start with quantity $\delta^{(2)}\gamma_{ab}=D_{\delta^{(2)}}\gamma_{ab}$, of which all we need is eqn(\ref{eq:47}):
\begin{align}
D_{\delta^{(2)}}\gamma_{ab}=D_{\delta^{(2)}}g(e_{a},e_{b})= & g(D_{\delta^{(2)}}e_{a},e_{b})+g(e_{a},D_{\delta^{(2)}}e_{b})\nonumber \\
= & 2\beta_{ab}^{(2)}\nonumber \\
= & 2K_{ab}{}^{i}\Psi_{i}-2(\mathcal{D}_{a}\Phi^{i})\cdot(\mathcal{D}_{b}\Phi_{i}) \nonumber \\
= & -2[\mathcal{D}_{a}(\mathcal{D}_{b}\Phi_{i})]\Phi^{i}.\label{eq:50}
\end{align}
We need also to obtain the quantity: 
\begin{equation}
 \delta^{(2)}K^{i}{}_{ab}=D_{\delta^{(2)}}K_{ab}{}^{i}
\end{equation}
or
\begin{align}
D_{\delta^{(2)}}K_{ab}{}^{i}=& -D_{\delta^{(2)}}g(D_{a}e_{b},n^{i}) \nonumber \\
= &  -g(D_{\delta^{(2)}}D_{a}e_{b},n^{i})-g(D_{a}e_{b},D_{\delta^{(2)}}n^{i}).\label{eq:51}
\end{align}
The first term on the right-hand side can be developed, using the
Ricci identity, as
\begin{widetext}
\begin{align}
-g(D_{\delta^{(2)}}D_{a}e_{b},n^{i})= & -g(R(\delta^{\mu(2)},e_{a})e_{b},n^{i})-g(D_{a}D_{\delta^{(2)}}e_{b},n^{i})\nonumber \\
= & -g(R(\delta^{\mu(2)},e_{a})e_{b},n^{i})-D_{a}\left[g(D_{\delta^{(2)}}e_{b},n^{i})\right]+g(D_{\delta^{(2)}}e_{b},D_{a}n^{i})\nonumber \\
= & -g(R(\delta^{\mu(2)},e_{a})e_{b},n^{i})-\nabla_{a}J_{b}^{(2)}{}^{i}+\beta_{bc}^{(2)}\cdot K_{a}{}^{ci}+\omega_{a}{}^{ij}\cdot J_{bj}^{(2)}\nonumber \\
= & -g(R(n_{j},e_{a})e_{b},n^{i})\Psi^{j}+g(R(e_{h},e_{a})e_{b},n^{i})J_{j}^{h}\Phi^{j}+K_{bc}{}^{k}\cdot K_{a}{}^{ci}\Psi_{k}\nonumber \\
 & -(\mathcal{D}_{b}\Phi^{l})\cdot(\mathcal{D}_{c}\Phi_{l})\cdot K_{a}{}^{ci}-[\mathcal{D}_{b}(\mathcal{D}_{c}\Phi_{o})]\cdot K_{a}{}^{ci}\Phi^{o}\nonumber \\
 & -\nabla_{a}J_{b}^{(2)}{}^{i}+\omega_{a}{}^{ij}\cdot J_{bj}^{(2)}\label{eq:52}
\end{align}
\end{widetext}
Using eq.(\ref{eq:43}), the second term on the right-hand side is
given by
\begin{equation}
-g(D_{a}e_{b},D_{\delta^{(2)}}n^{i})=\gamma_{ab}{}^{c}\cdot J_{c}^{(2)}{}^{i}.\label{eq:53}
\end{equation}
We find then that
\begin{widetext}
\begin{align}
D_{\delta^{(2)}}K_{ab}{}^{i}= & -\nabla_{a}J_{b}^{(2)}{}^{i}+\omega_{a}{}^{ij}\cdot J_{bj}^{(2)}+\gamma_{ab}{}^{c}\cdot J_{c}^{(2)}{}^{i}\nonumber \\
 & +K_{bc}{}^{k}\cdot K_{a}{}^{ci}\Psi_{k}-(\mathcal{D}_{b}\Phi^{l})\cdot(\mathcal{D}_{c}\Phi_{l})\cdot K_{a}{}^{ci}-[\mathcal{D}_{b}(\mathcal{D}_{c}\Phi_{o})]\cdot K_{a}{}^{ci}\Phi^{o}\nonumber \\
 & -g(R(n_{j},e_{a})e_{b},n^{i})\Psi^{j}+g(R(e_{h},e_{a})e_{b},n^{i})J_{j}^{h}\Phi^{j}\nonumber \\
= & -\mathcal{D}_{a}J_{b}^{(2)}{}^{i}+K_{bc}{}^{k}\cdot K_{a}{}^{ci}\Psi_{k}-g(R(n_{j},e_{a})e_{b},n^{i})\Psi^{j}\nonumber \\
 & -(\mathcal{D}_{b}\Phi^{l})\cdot(\mathcal{D}_{c}\Phi_{l})\cdot K_{a}{}^{ci}-[\mathcal{D}_{b}(\mathcal{D}_{c}\Phi_{o})]\cdot K_{a}{}^{ci}\Phi^{o}+g(R(e_{h},e_{a})e_{b},n^{i})J_{j}^{h}\Phi^{j}\nonumber \\
= & -\mathcal{D}_{a}\mathcal{D}_{b}\Psi^{i}-\mathcal{D}_{a}\left[K_{b}{}^{ic}(\mathcal{D}_{c}\Phi_{j})\Phi^{j}\right]+K_{bck}\cdot K_{a}{}^{ci}\Psi^{k}-g(R(n_{k},e_{a})e_{b},n^{i})\Psi^{k}\nonumber \\
 & -(\mathcal{D}_{b}\Phi^{l})\cdot(\mathcal{D}_{c}\Phi_{l})\cdot K_{a}{}^{ci}-[\mathcal{D}_{b}(\mathcal{D}_{c}\Phi_{o})]\cdot K_{a}{}^{ci}\Phi^{o}+g(R(e^{h},e_{a})e_{b},n^{i})\cdot\left(\mathcal{D}_{h}\Phi_{j}\right)\cdot\Phi^{j}\nonumber \\
= & -\mathcal{D}_{a}\mathcal{D}_{b}\Psi^{i}+K_{bck}\cdot K_{a}{}^{ci}\Psi^{k}-g(R(n_{k},e_{a})e_{b},n^{i})\Psi^{k}\nonumber \\
 & -\left[\mathcal{D}_{a}K_{b}{}^{ic}\right](\mathcal{D}_{c}\Phi_{j})\Phi^{j}-K_{b}{}^{ic}\left[\mathcal{D}_{a}(\mathcal{D}_{c}\Phi_{j})\right]\Phi^{j}-K_{b}{}^{ic}(\mathcal{D}_{c}\Phi_{j})\left(\mathcal{D}_{a}\Phi^{j}\right)\nonumber \\
 & -(\mathcal{D}_{b}\Phi^{l})\cdot(\mathcal{D}_{c}\Phi_{l})\cdot K_{a}{}^{ci}-[\mathcal{D}_{b}(\mathcal{D}_{c}\Phi_{o})]\cdot K_{a}{}^{ci}\Phi^{o}+g(R(e^{h},e_{a})e_{b},n^{i})\cdot\left(\mathcal{D}_{h}\Phi_{j}\right)\cdot\Phi^{j}\label{eq:6-15}
\end{align}
\end{widetext}
So, substituting the results into the eq.(\ref{eq:49}) we find the second order equation of motion in the form: 
\begin{equation}
 \triangle\Psi^{i}+K_{ck}{}^{a}\cdot K_{a}{}^{ci}\Psi^{k}+g(R(n_{k},e^{b})e_{b},n^{i})\Psi^{k}=\mathcal{F}^i\label{eq:6-19}
\end{equation}
where the source $\mathcal{F}^i$ is given in terms of the first order perturbations
\begin{align}
\mathcal{F}^i = & 2\left(\mathcal{D}_{a}\mathcal{D}_{b}\Phi^{i}\right)K^{ab}{}_{j}\Phi^{j}-2K^{ab}{}_{j}\Phi^{j}K_{bck}\Phi^{k}K^{ic}{}_{a}\nonumber \\
 & -2K^{ab}{}_{j}\Phi^{j}g\left(n^{i},R\left(n_{k},e_{a}\right)e_{b}\right)\Phi^{k}\label{eq:6-20}
\end{align}
It should be mentioned that only the mixed term of the equation (\ref{eq:49}) has contributed to the source $\mathcal{F}^i$.

Even when the background geometry is flat so that $R^{\alpha}_{\beta\gamma\delta}=0$, eq.(\ref{eq:6-19}) is extremely
complicated, involving scalars in the extrinsic geometry ($K_{ba}^i$ and $\omega_{a}^{ji}$) in combinations which, it appears, cannot be eliminated
in favor of intrinsic geometric scalars. If we can choose our normal vectors such that all but one of them, for example,
$n^{1}$, are parallel transported along any curve on the world sheet,
\begin{equation}
 D_a n^i=0,\label{eq:58}
\end{equation}
then $\omega_{a}^{ji}=0$ for all $i$ and $j$, because of its antisymmetry with respect to normal indices. In addition, 
the conditions eq.(\ref{eq:58}) imply that the only linear combination of extrinsic curvature tensors which is
nonvanishing is the one that corresponds to the exceptional normal direction:
\begin{equation}
 K_{ba}^i=0,\quad i=2,\ldots,p.\label{eq:59}
\end{equation}
These conditions are consistent with a reasonably large class of geometries. Any defect in Minkowski space which lies
in a $n$-dimensional plane will satisfy these conditions. A $n$-dimensional defect in de Sitter space with $p-1$ fixed meridians will
also satisfy eqs.(\ref{eq:59}). A less trivial example satisfying eqs.(\ref{eq:59}) is a string in Schwarzchild space
on a fixed meridian \cite{Gu93}.

Let us suppose, in addition, one more simplification which occurs whenever $p=2$, an example of which is provided by
a string in any four-dimensional manifold. For then, the coupling between the two scalar field components vanishes.
Equation (\ref{eq:6-19}) then reduces to the form of a Klein-Gordon equation:
\begin{equation}
 \square_\gamma\Psi^i +\mathcal{V}^i \Psi^i=\mathcal{F}^{i},\quad i=1,2 
\end{equation}
where $\square=\gamma^{ab}\nabla_{a}\nabla_{b}$ is the d’Alambertian, 
with
\begin{eqnarray}
 \mathcal{V}^{1}& = &K_{ac} K^{ac}+\gamma^{ab} y^{\mu}_{,a} y^{\nu}_{,b} R_{\mu\alpha\nu\beta} n^{\alpha}_{1} n^{\beta}_{1} \\
\mathcal{V}^{2}& = &\gamma^{ab} y^{\mu}_{,a} y^{\nu}_{,b} R_{\mu\alpha\nu\beta} n^{\alpha}_{2} n^{\beta}_{2}
 \end{eqnarray}
and the source $\mathcal{F}^{i}$: 

\begin{eqnarray}
 \mathcal{F}^{1} & = &2 K^{ab}{} \Phi^{1}\left(\nabla_{a}\nabla_{b}-K_{bc}K^{c}{}_{a}-y^{\mu}_{,a} y^{\nu}_{,b} R_{\mu\alpha\nu\beta} n^{\alpha}_{1} n^{\beta}_{1}\right)\Phi^{1} \nonumber \\
 \mathcal{F}^{2} & = & 0
 \end{eqnarray}
$\nabla_{a}$ is the strings world-sheet covariant derivative.

\section{Discussion}

In this paper we have studied membranes of arbitrary dimension moving
in any spacetime, using the perturbative scheme of Garriga and Vilenkin
\cite{GarVil91}. We adopted the kinematical framework of Capovilla and Guven
\cite{CapGuv95} in order to treat perturbations. Carter have also used kinematical terms in order to
study the dynamics of relativistic brane models \cite{Car01}. The physical measure of
the first and second order perturbations of $y^{\mu}$ is given
by two distinct scalar fields, which both live on the world sheet.
Our results for the equations of motion and the first order equations
of motion are in agreement with the independently obtained results
in a number of papers \cite{CapGuv95,Gu93,NicLar01}. The expression we derived for the
second order equations of motion is compact and can be seen as nontrivially
coupled scalar wave equations for a multiplet of scalar field, with
a variable mass that depends on a particular projection of the curvature
of spacetime and on the extrinsic geometry. It also has a source term which is given
in terms of the first order perturbations.

It is highly interesting to apply our result to strings moving in
curved spacetime in order to study, in a precise manner, the string
behavior near the black hole horizon, issues first raised by Susskind
\cite{Sus94}. Along these lines it has already been studied \cite{NicLar99}
an oscillating circular string in Schwarzchild background
to zeroth and first order. In this work it was calculated both the
radial and angular spreading of the string, as the string approaches
the black hole horizon. It was found that the radial spreading is
suppressed by the Lorentz contraction and the string appears (to an
asymptotic observer) as wrapping around the event horizon. We plan
to calculate and include the second order terms and thus analyze how
the string oscillators spread over the event horizon in a future paper.
Notice that the second order perturbations are necessary for a consistent
discussion of the energy. Hopefully we might understand the entropy
of the black hole in terms of string degrees of freedom.

\appendix*
\section{}
computation of $\gamma^{ab}D_{\delta^{(2)}}K_{ab}{}^{i}$:
\begin{widetext}
\begin{align}
\gamma^{ab}D_{\delta^{(2)}}K_{ab}{}^{i}= & \gamma^{ab}\left[-\mathcal{D}_{a}\mathcal{D}_{b}\Psi^{i}+K_{bck}\cdot K_{a}{}^{ci}\Psi^{k}-g(R(n_{k},e_{a})e_{b},n^{i})\Psi^{k}\right]+\nonumber \\
 & \gamma^{ab}\left[-\left[\mathcal{D}_{a}K_{b}{}^{ic}\right](\mathcal{D}_{c}\Phi_{j})\Phi^{j}-K_{b}{}^{ic}\left[\mathcal{D}_{a}(\mathcal{D}_{c}\Phi_{j})\right]\Phi^{j}-K_{b}{}^{ic}(\mathcal{D}_{c}\Phi_{j})\left(\mathcal{D}_{a}\Phi^{j}\right)\right]\nonumber \\
 & \gamma^{ab}\left[-(\mathcal{D}_{b}\Phi^{l})\cdot(\mathcal{D}_{c}\Phi_{l})\cdot K_{a}{}^{ci}-[\mathcal{D}_{b}(\mathcal{D}_{c}\Phi_{o})]\cdot K_{a}{}^{ci}\Phi^{o}+g(R(e^{h},e_{a})e_{b},n^{i})\cdot\left(\mathcal{D}_{h}\Phi_{j}\right)\cdot\Phi^{j}\right]\nonumber \\
= & -\widetilde{\triangle}\Psi^{i}+K_{ck}{}^{a}\cdot K_{a}{}^{ci}\Psi^{k}-g(R(n_{k},e^{b})e_{b},n^{i})\Psi^{k}\nonumber \\
 & -\left[\mathcal{D}_{a}K{}^{aic}\right](\mathcal{D}_{c}\Phi_{j})\Phi^{j}-K{}^{aic}\left[\mathcal{D}_{a}(\mathcal{D}_{c}\Phi_{j})\right]\Phi^{j}-K{}^{aic}(\mathcal{D}_{c}\Phi_{j})\left(\mathcal{D}_{a}\Phi^{j}\right)\nonumber \\
 & -(\mathcal{D}_{b}\Phi^{l})\cdot(\mathcal{D}_{c}\Phi_{l})\cdot K{}^{bci}-[\mathcal{D}_{b}(\mathcal{D}_{c}\Phi_{o})]\cdot K{}^{bci}\Phi^{o}+g(R(e^{h},e^{b})e_{b},n^{i})\cdot\left(\mathcal{D}_{h}\Phi_{j}\right)\cdot\Phi^{j}\nonumber \\
= & -\widetilde{\triangle}\Psi^{i}+K_{ck}{}^{a}\cdot K_{a}{}^{ci}\Psi^{k}-g(R(n_{k},e^{b})e_{b},n^{i})\Psi^{k}\nonumber \\
 & -2K{}^{aic}(\mathcal{D}_{c}\Phi_{j})\left(\mathcal{D}_{a}\Phi^{j}\right)-2K{}^{aic}\left[\mathcal{D}_{a}(\mathcal{D}_{c}\Phi_{j})\right]\Phi^{j}\nonumber \\
 & -\left[\mathcal{D}^{c}K_{b}{}^{bi}\right](\mathcal{D}_{c}\Phi_{j})\Phi^{j}\label{eq:6-16}
\end{align}
computation of $\left(D_{\delta^{(2)}}\gamma^{ab}\right)K_{ab}{}^{i}$:
\begin{align}
\left(D_{\delta^{(2)}}\gamma^{ab}\right)K_{ab}{}^{i}= & \left[-2K^{ab}{}_{i}\Psi^{i}+2(\mathcal{D}^{a}\Phi^{i})\cdot(\mathcal{D}^{b}\Phi_{i})+2[\mathcal{D}^{a}(\mathcal{D}^{b}\Phi_{i})]\Phi^{i}\right]K_{ab}{}^{i}\nonumber \\
= & -2K^{ab}{}_{j}\cdot K_{ab}{}^{i}\Psi^{j}+2(\mathcal{D}^{a}\Phi^{j})\cdot(\mathcal{D}^{b}\Phi_{j})K_{ab}{}^{i}+2[\mathcal{D}^{a}(\mathcal{D}^{b}\Phi_{j})]K_{ab}{}^{i}\Phi^{j}\label{eq:6-17}
\end{align}
 computation of $\left(D_{\delta}\gamma^{ab}\right)\left(D_{\delta}K_{ab}{}^{i}\right)$:
\begin{align}
\left(D_{\delta}\gamma^{ab}\right)\left(D_{\delta}K_{ab}{}^{i}\right)= & \left[-2K^{ab}{}_{j}\Phi^{j}\right]\left[-\mathcal{D}_{a}\mathcal{D}_{b}\Phi^{i}+\left[g\left(n^{i},R\left(n_{k},e_{a}\right)e_{b}\right)+K_{bck}\cdot K^{ic}{}_{a}\right]\Phi^{k}\right]\nonumber \\
= & 2\mathcal{D}_{a}\mathcal{D}_{b}\Phi^{i}K^{ab}{}_{j}\Phi^{j}-2K^{ab}{}_{j}\Phi^{j}g\left(n^{i},R\left(n_{k},e_{a}\right)e_{b}\right)\Phi^{k}\nonumber \\
 & -2K^{ab}{}_{j}\Phi^{j}K_{bck}\cdot K^{ic}{}_{a}\Phi^{k}\label{eq:6-18}
\end{align}
\end{widetext}
\bibliography{sop}

\providecommand{\noopsort}[1]{}\providecommand{\singleletter}[1]{#1}%
\begin{thebibliography}{10}

\bibitem{CapGuv95}
R.~Capovilla and J.~Guven.
\newblock {\em Phys.\ Rev. D}, 51:6736, 1995.

\bibitem{Car93}
B.~Carter.
\newblock {\em Phys.\ Rev. D}, 48:4835, 1993.

\bibitem{Car01}
B.~Carter.
\newblock {\em Int. J. Theor. \ Phys.}, 40:2099, 2001.

\bibitem{deVSan87}
H.~J. de~Vega and N.~Sanchez.
\newblock {\em Phys.\ Lett. B}, 197:320, 1987.

\bibitem{FrLa94}
V.~P. Frolov and A.~L. Larsen.
\newblock {\em Nucl.\ Phys. B}, 414:129, 1994.

\bibitem{GarVil91}
J.~Garriga and A.~Vilenkin.
\newblock {\em Phys.\ Rev. D}, 44:1007, 1991.

\bibitem{Gu93}
J.~Guven.
\newblock {\em Phys.\ Rev. D}, 48:5562, 1993.

\bibitem{HawEl73}
S.~W. Hawking and G.~F.~R. Ellis.
\newblock {\em The large scale structure of space-time}.
\newblock Cambridge University Press, 1973.

\bibitem{NicLar99}
A.~L. Larsen and A.~Nicolaidis.
\newblock {\em Phys.\ Rev. D}, 60:024012, 1999.

\bibitem{NicLar01}
A.~L. Larsen and A.~Nicolaidis.
\newblock {\em Phys.\ Rev. D}, 63:125006, 2001.

\bibitem{Sus94}
L.~Susskind.
\newblock {\em Phys.\ Rev. D}, 49:6606, 1994.

\end{thebibliography}

\end{document}